\documentclass[prl,twocolumn,superscriptaddress,nofootinbib,amsmath,amssymb,longbibliography]{revtex4-1}
\usepackage{graphicx}
\usepackage[usenames,dvipsnames]{xcolor}
\usepackage[colorlinks=true,citecolor=Blue,linkcolor=RubineRed,urlcolor=Blue]{hyperref}
\begin{document}

\title{Kibble-Zurek Mechanism for Nonequilibrium Phase Transitions in Driven Systems with Quenched Disorder}
 
\author{C. J. O. Reichhardt}
\affiliation{Theoretical Division and Center for Nonlinear Studies,
Los Alamos National Laboratory, Los Alamos, New Mexico 87545, USA}
\author{A. del Campo}
\affiliation{Department  of  Physics  and  Materials  Science,  University  of  Luxembourg,  L-1511  Luxembourg, Luxembourg}
\affiliation{Donostia International Physics Center,  E-20018 San Sebasti\'an, Spain}
\author{C. Reichhardt}
\affiliation{Theoretical Division and Center for Nonlinear Studies,
Los Alamos National Laboratory, Los Alamos, New Mexico 87545, USA}

\date{\today}

\begin{abstract}
We numerically study the density of topological defects for a two-dimensional assembly of particles driven over quenched disorder as a function of quench rate through the nonequilibrium phase transition from a plastic disordered flowing state to a moving anisotropic crystal. A dynamical ordering transition of this type occurs for vortices in type-II superconductors, colloids, and other particle-like systems in the presence of random disorder. We find that on the ordered side of the transition, the density of topological defects $\rho_d$ scales as a power law, $\rho_d \propto 1/t_{q}^\beta$, where $t_{q}$ is the time duration of the quench across the transition. This type of scaling is predicted in the Kibble-Zurek mechanism for varied quench rates across a continuous phase transition. We show that scaling with the same exponent holds for varied strengths of quenched disorder. The value of the exponent can be connected to the directed percolation universality class. Our results suggest that the Kibble-Zurek mechanism can be applied to general nonequilibrium phase transitions.   
\end{abstract}

\maketitle

\section{Introduction}
In systems that exhibit second-order phase transitions, a disordered phase
on one side of the transition, such as a liquid or glass state,
can be characterized by the presence of topological defects,
while on the other side of the transition,
there is an ordered or defect-free state such as a crystal.
The defect-free ordered phase arises
when the parameter controlling the transition
is changed very slowly, so that the system
remains in the adiabatic limit.
If the rate of change through
the transition increases,
a portion of the topological defects do not have time to 
annihilate, and persist on the ordered side of the transition. 
A scenario for understanding the
behavior for varied quench rates across a continuous phase transition 
is the Kibble-Zurek (KZ) mechanism,  
which predicts a universal power law for the defect density
$\rho_{d} \propto t_{q}^{-\beta}$,
where $t_{q}$ is the time duration of the quench
through the transition,
so that slower quenches produce
fewer defects 
\cite{Kibble76,Zurek85,Zurek96,delCampo14}. 
The KZ mechanism has been studied in
a variety of systems at the transition into an ordered phase, 
where the scaling exponents can be related to
the universality class of the underlying 
phase transition
\cite{Bowick94,Weiler08,Ulm13,Pyka13,Deutschlander15,Kessling19,Ko19}.   

The KZ scenario has generally been applied to systems
that have equilibrium phase transitions; 
however, there have been recent proposals to
use the KZ scenario to
address transitions between different nonequilibrium steady states
\cite{Ducci99,Casado01,Casado06,Casado07,Miranda12,Miranda13},
such as those that occur in optical systems or 
for Rayleigh-B{\`e}nard convection,
where defects can arise in otherwise hexagonal ordered lattices. 
Another class of nonequilibrium  
systems
consist of assemblies of particles
driven over quenched disorder that also exhibit behavior
consistent with a continuous phase transition from a disordered state to
an ordered state
\cite{Bhattacharya93,Koshelev94,Bohlein12a,Reichhardt17,Sandor17a} or
from a dynamical fluctuating
state to a non-fluctuating state
\cite{Hinrichsen00,Takeuchi07,Corte08,Fruchart21}.
An open question is whether 
the KZ scenario could also apply to nonequilibrium phase transitions
for varied sweep rates through the transition.
Nonequilibrium systems have several features that could
make them ideal for studying the KZ scenario.
They often contain very well defined topological defects,
and the transition can occur at $T = 0$, 
so that thermal effects such as critical coarsening
on the ordered side of the transition are absent.   

One of the best examples of a system that shows evidence for
a continuous nonequilibrium phase transition as a function of
a continuously changing driving parameter
from a disordered fluctuating state with
a high density of topological defects
to a dynamically ordered nonfluctuating state in which
topological defects are scarce or absent
is superconducting vortices driven over quenched disorder
\cite{Bhattacharya93,Koshelev94,Yaron95,Hellerqvist96,Moon96,Ryu96,Balents98,LeDoussal98,Pardo98,Olson98a,Kolton99,Troyanovski99,Xiao00,Okuma08}.
At $T = 0$ and in the absence of quenched disorder,
a superconducting vortex system  
forms a triangular lattice free of defects; however,
when quenched disorder is present,
a disordered state containing numerous
topological defects can appear that can be characterized
as a vortex glass \cite{Blatter94}.
A finite external drive in the form of an applied current
causes
vortices in the disordered state to
depin and move \cite{Reichhardt17,Blatter94}.
For drives above depinning,
the system enters a strongly fluctuating or plastically flowing state
in which there is a fluctuating number of topological defects,
while at higher drives,
there is a critical driving force above which the vortices
dynamically order
into a state with zero or a small number of
topological defects
\cite{Koshelev94,Reichhardt17,Moon96,Ryu96,Balents98,LeDoussal98,Pardo98,Olson98a}. 
This ordered state is not isotropic but takes the form of a moving anisotropic
crystal \cite{Reichhardt17,Ryu96,Balents98,Pardo98,Olson98a}
or a moving smectic state
\cite{Moon96,Balents98,LeDoussal98,Pardo98,Olson98a,Kolton99}.
Here,
a small number of topological defects in the form of dislocations
are present which have their Burgers 
vectors aligned in the direction of the drive.
The critical drive amplitude at which
the ordering transition occurs is a function of vortex
density and disorder strength \cite{Reichhardt17}.
The ordering transition
has been observed experimentally by direct imaging
\cite{Pardo98,Troyanovski99}, neutron scattering \cite{Yaron95}, 
changes in the noise \cite{Olson98a,Kolton99,Okuma08},
and changes in transport curve features 
\cite{Bhattacharya93,Reichhardt17,Hellerqvist96,Moon96,Ryu96,Olson98a,Xiao00}. 
This same type of dynamical ordering transition
is general to the broader class of particle-like assemblies driven
over quenched disorder,
and it has been studied for colloidal 
particles moving over disordered landscapes
\cite{Reichhardt17,Chen11,Granato11,Tierno12a,Bohlein12a},
moving Wigner crystals \cite{Reichhardt01},
sliding charged systems \cite{Danneau02},
driven pattern forming states \cite{Sengupta10,Reichhardt03a},
driven dislocations \cite{Zhou15},
and magnetic skyrmions moving over quenched disorder \cite{Reichhardt15}.
Since the transition from 
the disordered state to the ordered state
is controlled by changing the driving amplitude,
the number of topological defects
on the ordered side of the transition
can be measured for varied drive sweep rates
across the transition to test the
predictions of the KZ scenario.  

Here, we numerically examine the density of defects
in driven vortex and colloidal systems at $T = 0$ 
for varied driving sweep rates
through the dynamical ordering transition.
We find that the defect density obeys
$\rho_{d} \propto \tau_{q}^\beta$, 
where $t_{q}$ is the time
required to go through the transition, consistent with the KZ scenario.
The same behavior appears
for various values of the quenched disorder and for both vortices and
colloids moving over a random substrate.
The exponent we find in both systems is
$\beta \approx 0.39$, which is consistent
with an underlying transition that falls in a directed percolation (DP)
universality 
class, which often describes
nonequilibrium phase transitions
from fluctuating states to non-fluctuating states
\cite{Hinrichsen00,Takeuchi07,Fruchart21}.  
These results could be tested experimentally
in superconducting vortex systems, colloids, 
or other driven systems with quenched disorder.  
Our results also imply that the KZ mechanism could be applied generally
to nonequilibrium phase transitions in the same way that it has
been applied to equilibrium phase transitions,
and that it could be tested in many other types of driven systems that 
show dynamical continuous transitions from disordered to ordered states.

\section{Results}
\noindent{\textbf{\textsf{Modeling and characterization of the nonequilibrium phase transition}}}\\ 
We use a well-established simulation model for
superconducting vortices driven over random disorder.
The vortices are represented as repulsively interacting point 
particles
which undergo
a dynamical ordering transition from 
a disordered plastic flow phase to a moving ordered crystal or
smectic state \cite{Koshelev94,Reichhardt17,Moon96,Ryu96,Olson98a,Kolton99}. 
We consider a two-dimensional (2D) system
of size $L \times L$, with $L=36\lambda$ in units of the London
penetration depth, containing a fixed number $N$ of
vortices produced by an external magnetic field ${\bf B}=B{\hat {\bf z}}$.
The vortex density is $n_v=N/L^2$.
The motion 
of vortex $i$ at position ${\bf R}_i$ is obtained by
integrating the overdamped equation of motion
\begin{equation}  
\eta \frac{d{\bf R}_{i}}{dt}= {\bf F}_{vv} + {\bf F}^{pp}_{i} + {\bf F}_{d} ,
\end{equation}
where the damping constant $\eta = 1.0$. 
The vortex-vortex interaction force 
${\bf F}^{vv}_{i} = \sum^{N}_{j=1}f_{0}K_{1}(R_{ij}/\lambda){\hat {\bf R}_{ij}}$, 
where $K_{1}$ is the modified Bessel function, $R_{ij}=|{\bf R}_i-{\bf R}_j|$ is the distance
between vortex $i$ and $i$, 
${\hat {\bf R}}_{ij}=({\bf R}_i-{\bf R}_j)/R_{ij}$,
$f_{0} = \phi^{2}_{0}/2\pi\mu_{0}\lambda^3$,
$\mu_0$ is the permeability of free space,
and $\phi_{0} = h/2e$ is the elementary flux quantum. 
The modified Bessel function $K_1$ falls off exponentially 
for $R_{ij}/\lambda > 1.0$.
The vortices are confined to the $x,y$ plane and the magnetic field
is applied in the $z$-direction.
The
quenched disorder is modeled by $N_p$ randomly placed
harmonically attractive pinning sites of 
radius $r_{p}$ with 
${\bf F}^{pp}_{i} = -\sum_{k=1}^{N_p}(F_{p}/r_{p})({\bf R}_{i} - {\bf R}^{(p)}_{k})\Theta(r_{p} - |{\bf R}_{i} - {\bf R}^{(p)}_{k}|)\hat {\bf R}_{ik}^{(p)}$
where $F_{p}$ is the maximum pinning force,
${\bf R}^{(p)}_{k}$ is the location of pinning site $k$,
and ${\hat {\bf R}}_{ik}^{(p)}=({\bf R}_i-{\bf R}_k^{(p)})/|{\bf R}_i-{\bf R}_k^{(p)}|$.
The pinning density is $n_p=N_p/L^2$.
The driving force ${\bf F}_{d}$ is applied uniformly to all the vortices,
and corresponds experimentally to a Lorentz
force ${\bf F}_{d} = {\bf B}\times {\bf J}=F_D f_0 {\hat {\bf x}}$, where ${\bf J}$ is
the applied current.  
The initial vortex configurations
are obtained using simulated annealing. We increase
the magnitude of the dimensionless driving force
from $F_D=0$ to a final maximum value
over a total time of $t_{q}$
in increments of
$\Delta F_D = 0.002$.
Time is reported in dimensionless units
achieved by scaling time by $\tau=\eta/f_0$.
We characterize the system 
by measuring the average velocity per vortex in the driving direction,
$\langle V\rangle = N^{-1}\sum_i^N {\bf v}_i \cdot {\bf \hat x}$ and
the fraction of sixfold coordinated vortices
$P_6=N^{-1}\sum_{i}^{N}\delta(z_i-6)$, where $z_i$ is the coordination number
of vortex $i$ obtained
from a Voronoi construction. 
We also compute the fraction of topological defects $P_{D}=1-P_6$.

\begin{figure}
\includegraphics[width=3.5in]{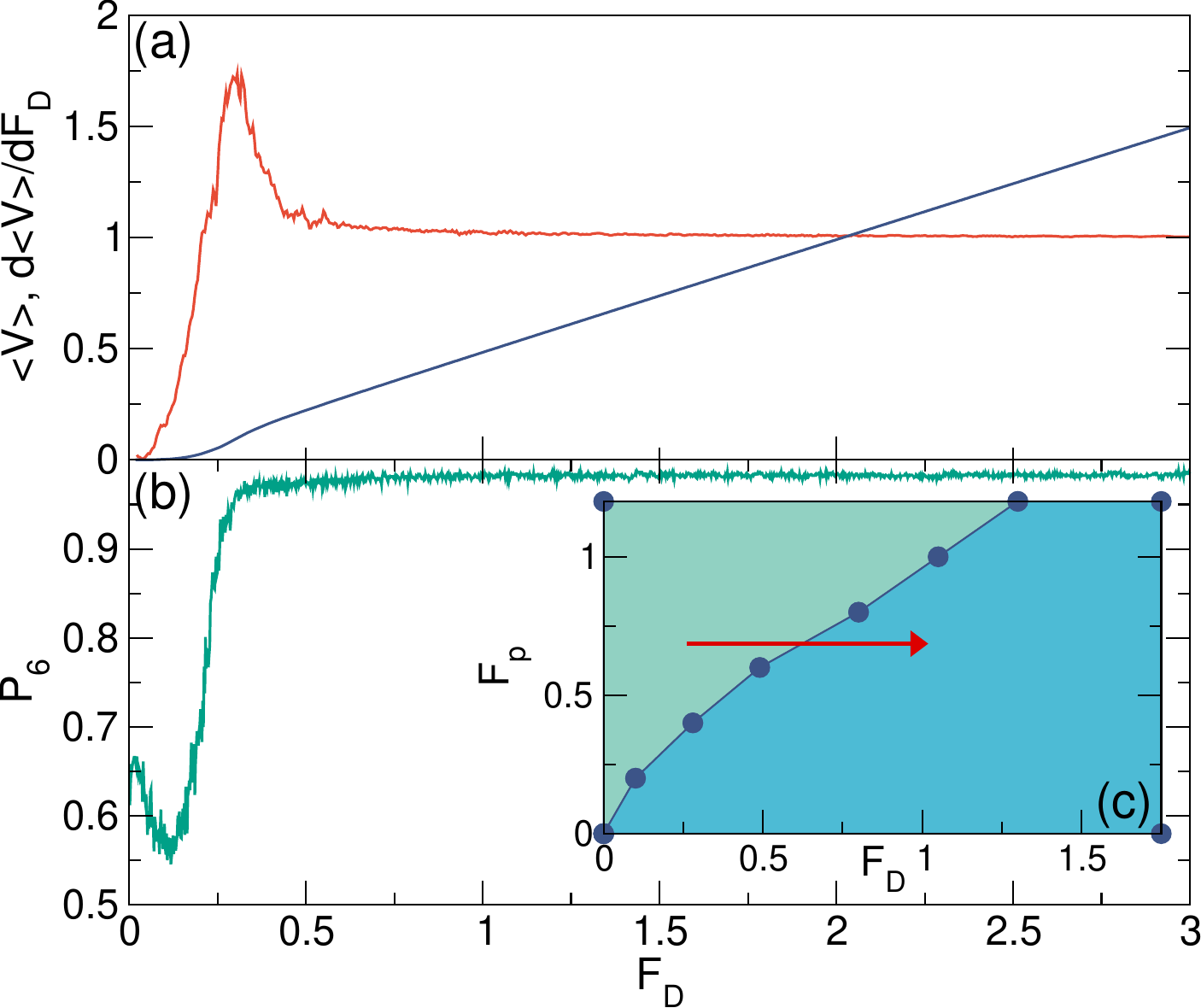}
\caption{ {\bf Dynamical phase transition in a superconducting vortex system.}
  (a) The velocity $\langle V\rangle$ versus applied
drive $F_{D}$ (blue) and the corresponding
$d\langle V\rangle/dF_D$ vs $F_D$ (red)  
for a 2D superconducting vortex system with quenched disorder
at vortex density $n_{v} = 1.0$, pinning density $n_{p} = 0.5$,
pinning force $F_{p} = 0.4$, and pinning radius $r_{p} = 0.3$.
The curves are obtained for a quench time of
$t_{q} = 7.5 \times 10^6$, which is considered the adiabatic limit.    
(b) The corresponding fraction of sixfold coordinated vortices
$P_{6}$ vs $F_{D}$.
A nonequilibrium transition from a disordered state
to an ordered state occurs for $F_{D} \approx 0.3$. 
(c) Dynamical phase diagram as a function of $F_p$ versus $F_D$ for the
same system
highlighting the disordered phase (green)
and the dynamically ordered phase (blue).
The arrow indicates the direction in which the transition is crossed
at different sweep rates.
}
\label{fig:1}
\end{figure}

\begin{figure}
\includegraphics[width=3.5in]{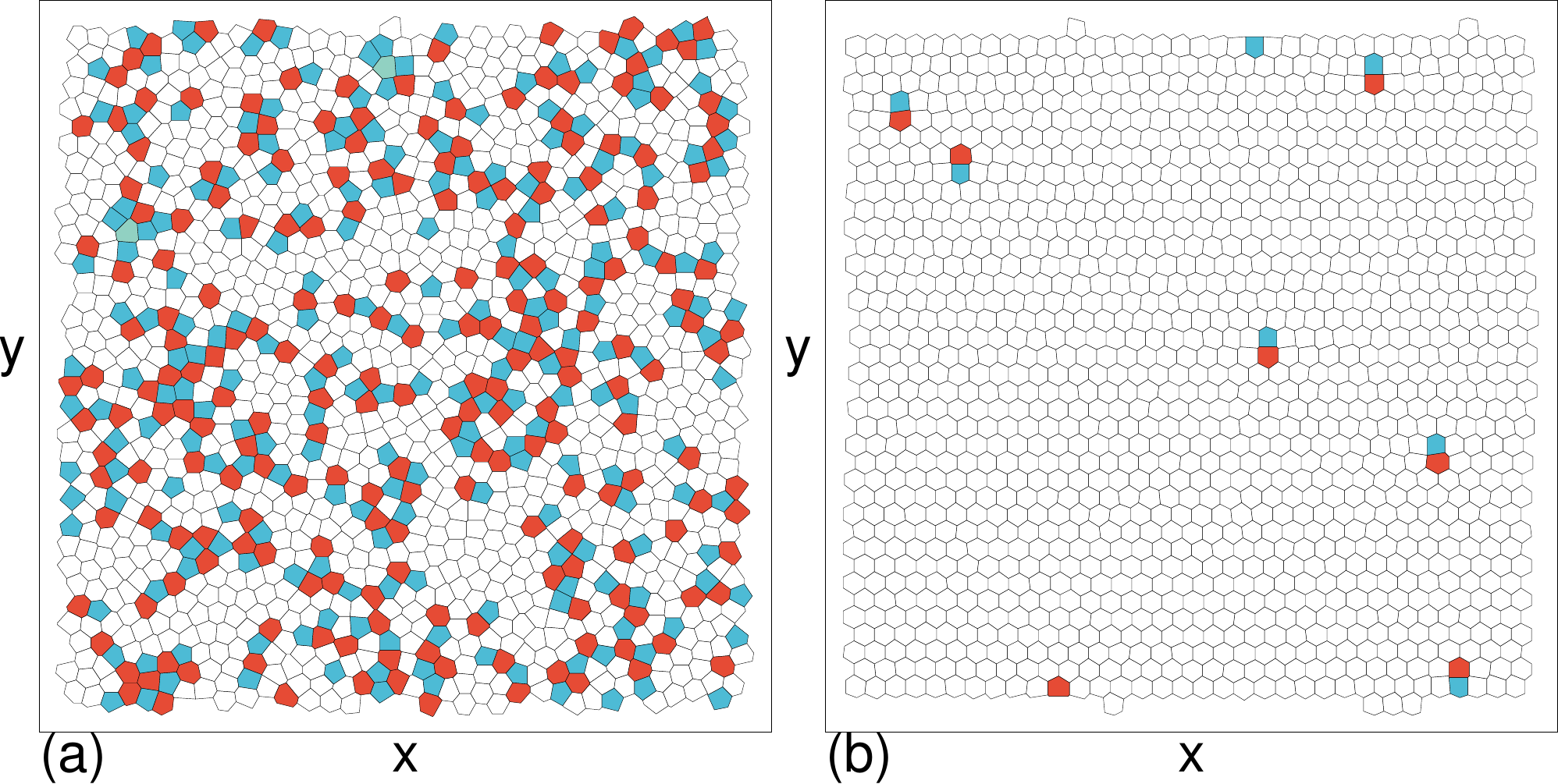}
\caption{{\bf Vortex configurations on either side of the dynamical phase transition.}
  Voronoi constructions of snapshots of the vortex positions
in the system from Fig.~\ref{fig:1}
with $n_v=1.0$, $n_p=0.5$, $F_p=0.4$, $r_p=0.3$, and $t_q=7.5 \times 10^6$.
White polygons are sixfold coordinated, red polygons are sevenfold
coordinated, and blue polygons are fivefold coordinated.
(a) $F_{D} = 0.1$ in the disordered phase with numerous defects.
(b) $F_{D} = 1.5$ in the dynamically ordered phase.
The vortex lattice is
aligned in the driving direction
and there is
a small number of aligned dislocations.
}
\label{fig:2}
\end{figure}

\noindent{\textbf{\textsf{Numerical results}}}\\ 
Figure~\ref{fig:1} illustrates the effect of changing $F_D$ in the
adiabatic limit for a system
with $n_{v} = 1.0$, $n_{p} = 0.5$, $F_{p} = 0.4$, and $r_{p} = 0.3$.
In Fig.~\ref{fig:1}(a) we plot the average vortex velocity
$\langle V\rangle$ along with $d\langle V\rangle/dF_D$ versus
$F_{D}$
for $t_{q} = 7.5 \times 10^6$. If $t_q$ is increased to a larger value,
the curves remain nearly unchanged.
Figure~\ref{fig:1}(b) shows the corresponding
fraction of sixfold coordinated vortices
$P_{6}$ versus $F_D$, where we would have
$P_6=1$ for a triangular lattice.
A depinning transition occurs near $F_{D} = 0.06$,
and for $F_{D} < 0.225 $ we find $P_{6} \approx 0.64$,
indicating a highly defected or disordered
state in which the vortices are undergoing plastic flow.
In Fig.~\ref{fig:2}(a) we show a
Voronoi plot of a snapshot
of the vortex positions from the system in
Fig.~\ref{fig:1} at $F_{D} = 0.1$, where numerous dislocations are present.
As Fig.~\ref{fig:1}(b) indicates, for $F_{D} > 0.4$ there is a regime
in which $P_{6}$ increases until it reaches the value
$P_{6} = 0.98$, where the vortices form an
anisotropic triangular lattice of the type
shown in Fig.~\ref{fig:2}(b) at $F_{D} = 1.5$ in the ordered state.
Here the lattice 
is aligned in the driving direction and contains a small number of
aligned dislocations,
forming a smectic state as studied previously
\cite{Moon96,Ryu96,Balents98,Pardo98,Olson98a}.
The critical force for the transition from the disordered state to
the dynamically ordered phase is
$F_{D}^c \approx 0.3$.
We do not observe any hysteresis
across the transition, which
has features consistent with a second-order transition.
In Fig.~\ref{fig:1}(a), there is
a peak in $d\langle V\rangle/dF_{D}$
in the plastic flow phase 
near $F_{D} = 0.15$ followed 
by a saturation near $F_{D} = 0.3$ to
$d\langle V\rangle/dF_D \approx 1.0$, in agreement with
previous studies of the
dynamical ordering transition
\cite{Bhattacharya93,Hellerqvist96,Ryu96,Kolton99,Xiao00,Okuma08}.   
The critical driving force for dynamic ordering also depends 
on the strength of the disorder,
as shown in Fig.~\ref{fig:1}(c) where we plot a dynamical phase diagram 
for the system in Fig.~\ref{fig:1}(a,b)
as a function of $F_{p}$ versus $F_{D}$ at $t_{q} = 7.5 \times 10^6$.
The disordered and ordered phases are highlighted, and
the arrow illustrates the direction in which we sweep across
the transition
at different rates.   

\begin{figure}
\includegraphics[width=3.5in]{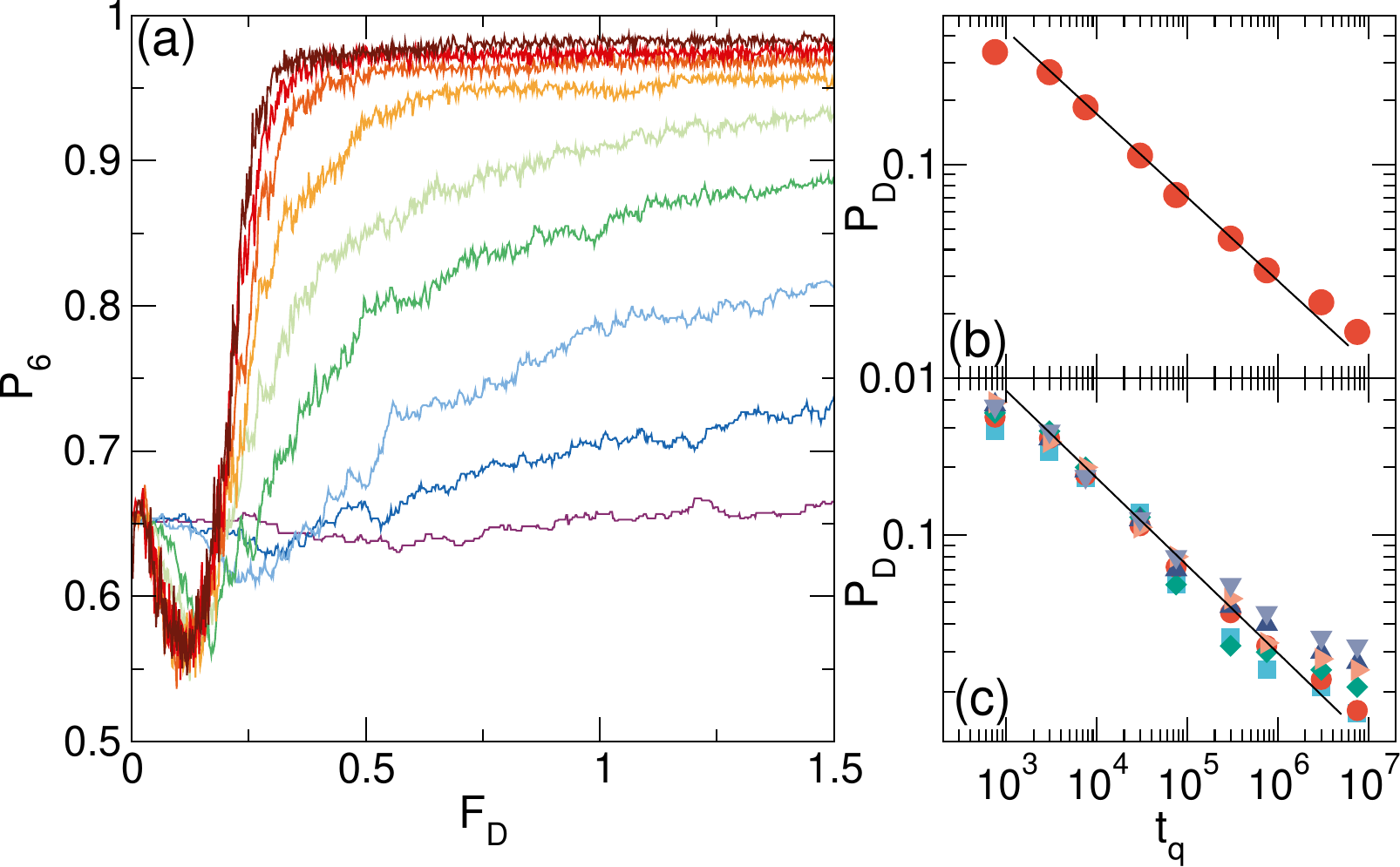}
\caption{
  {\bf Transition from the disordered to the ordered state
    for superconducting vortices as a function
    of quenching speed.}
(a) $P_{6}$ versus $F_{D}$
for the system in Fig.~\ref{fig:1} with $n_v=1.0$, $n_p=0.5$, $F_p=0.4$,
and $r_p=0.3$ at
$t_{q} = 7.5\times 10^6$, $3\times 10^6$, $7.5\times 10^5$, $3\times 10^5$,
$7.5\times 10^4$, $3\times 10^4$, $7.5 \times 10^3$, $3\times 10^3$,
and $750$,
from top to bottom.
(b) The final defect density $P_{D} = 1 -P_{6}$
at $F_D=1.5$ in the same system versus $t_{q}$.
The solid line is a power-law fit with $\beta = -0.385$. 
(c)
$P_{D}$ versus $t_{q}$ for the same system
at different values of
$F_{p} = 0.2$ (blue squares),
$0.4$ (red circles),
$0.6$ (green diamonds),
$0.8$ (blue up triangles),
$1.0$ (orange right triangles),
and $1.2$ (purple down triangles). The solid line
is a power law fit with $\beta = -0.39$.
}
\label{fig:3}
\end{figure}

Now that we have established the driving force which separates the
disordered and ordered phases, we can cross this transition
for varied $t_{q}$.
In Fig.~\ref{fig:3}(a) we plot $P_{6}$ versus $F_{D}$ 
for the system in Fig.~\ref{fig:1} at 
$t_{q} = 7.5\times 10^6$, $3\times 10^6$, $7.5\times 10^5$,
$3\times 10^5$, $7.5\times 10^4$,
$3\times 10^4$, $7.5 \times 10^3$, $3\times 10^3$, and $750$, 
showing that as the quench rate increases and $t_q$ becomes smaller,
the fraction of six-fold coordinated particles on the ordered side of the
critical transition value $F^{c}_{D} = 0.3$ drops. 

\begin{figure}
\includegraphics[width=3.5in]{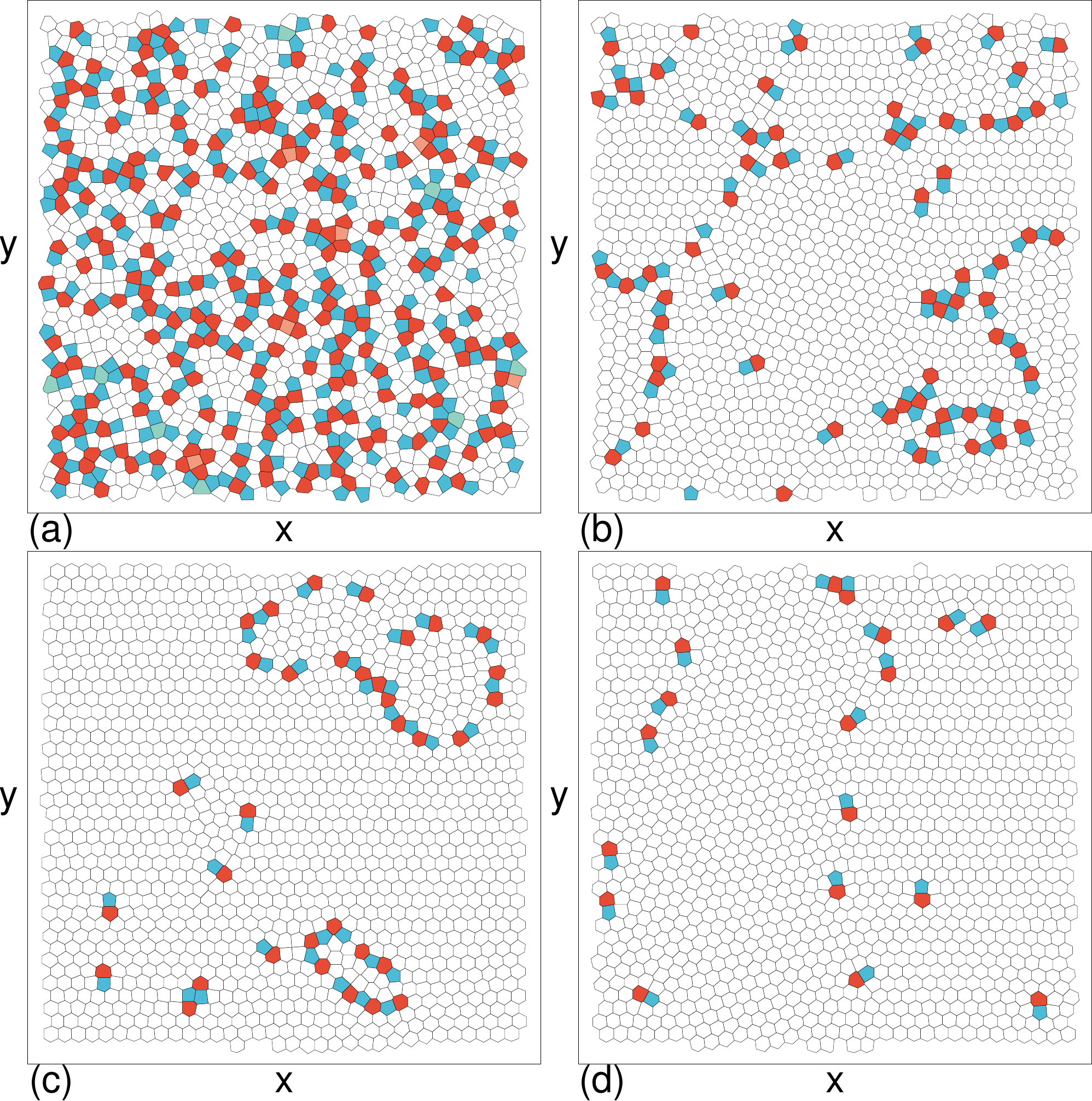}
\caption{
  {\bf Final ordering of lattice at different quenching speeds.}
Voronoi plots of the
final $F_D=1.5$ state for the system in Fig.~\ref{fig:3}(a,b)
with $n_v=1.0$, $n_p=0.5$, $F_p=0.4$, and $r_p=0.3$ at
(a) $t_{q} = 7.5\times 10^3$,
(b) $3 \times 10^4$,
(c) $7.5 \times 10^4$, and
(d) $3\times 10^5$.
The number of defects is larger for smaller $t_{q}$ or a faster sweep rate.
}
\label{fig:4}
\end{figure}

We illustrate Voronoi plots
of the final $F_D=1.5$ state
for the system in Fig.~\ref{fig:3} at 
$t_{q} = 7.5\times 10^3$ in Fig.~\ref{fig:4}(a),
$t_q=3\times 10^4$ in Fig.~\ref{fig:4}(b),
$t_q=7.5\times 10^4$ in Fig.~\ref{fig:4}(c),
and $3\times 10^5$ in Fig.~\ref{fig:4}(d),
indicating that the number of
remaining defects is larger for smaller $t_{q}$. 
Figure~\ref{fig:2}(b) shows the same sample in the adiabatic limit with
$t_{q} = 7.5 \times 10^6$. 
The defects that appear on the ordered side of the transition 
generally take the form of
dislocations composed of pairs of fivefold and sevenfold coordinated particles.
At larger $t_{q}$, the dislocations have their Burgers vectors
oriented in the direction of drive,
giving rise to a smectic ordering in which
the system can be regarded as a set of one-dimensional (1D) moving channels. 
In Fig.~\ref{fig:3}(b) we plot $P_D=1-P_6$, the density of topological
defects at the final $F_D=1.5$ state on the ordered side of the transition,
versus $t_{q}$.
We find $P_{D} \propto t_{q}^\beta$, 
where the solid line indicates a fit with $\beta = -0.385$.
The KZ scenario predicts a power law decay
of $P_D$ with increasing $t_{q}$.
If we choose a final value of $F_D$ which is closer to but still above
the critical transition drive $F_c$,
the scaling is not as good at smaller $t_{q}$
but we still find the same exponent
for the larger values of $t_{q}$.     

The plots of $P_D$ versus $t_q$
in Fig~\ref{fig:3}(c)
indicate that the same
scaling behavior
remains robust over a range of different values
of $F_{p}$ from $F_p= 0.2$ to $F_p=1.2$.
Since $F_c$ varies with $F_p$,
for each sample we select a final value of $F_D$ that
is the same distance above
$F_{c}$ for consistency.
The straight line is a power-law fit
with $\beta = 0.39$.
For large $F_{p}$, the scaling at larger values of
$t_{q}$ is not as good because
more defects become trapped by the pinning.

\begin{figure}
\includegraphics[width=3.5in]{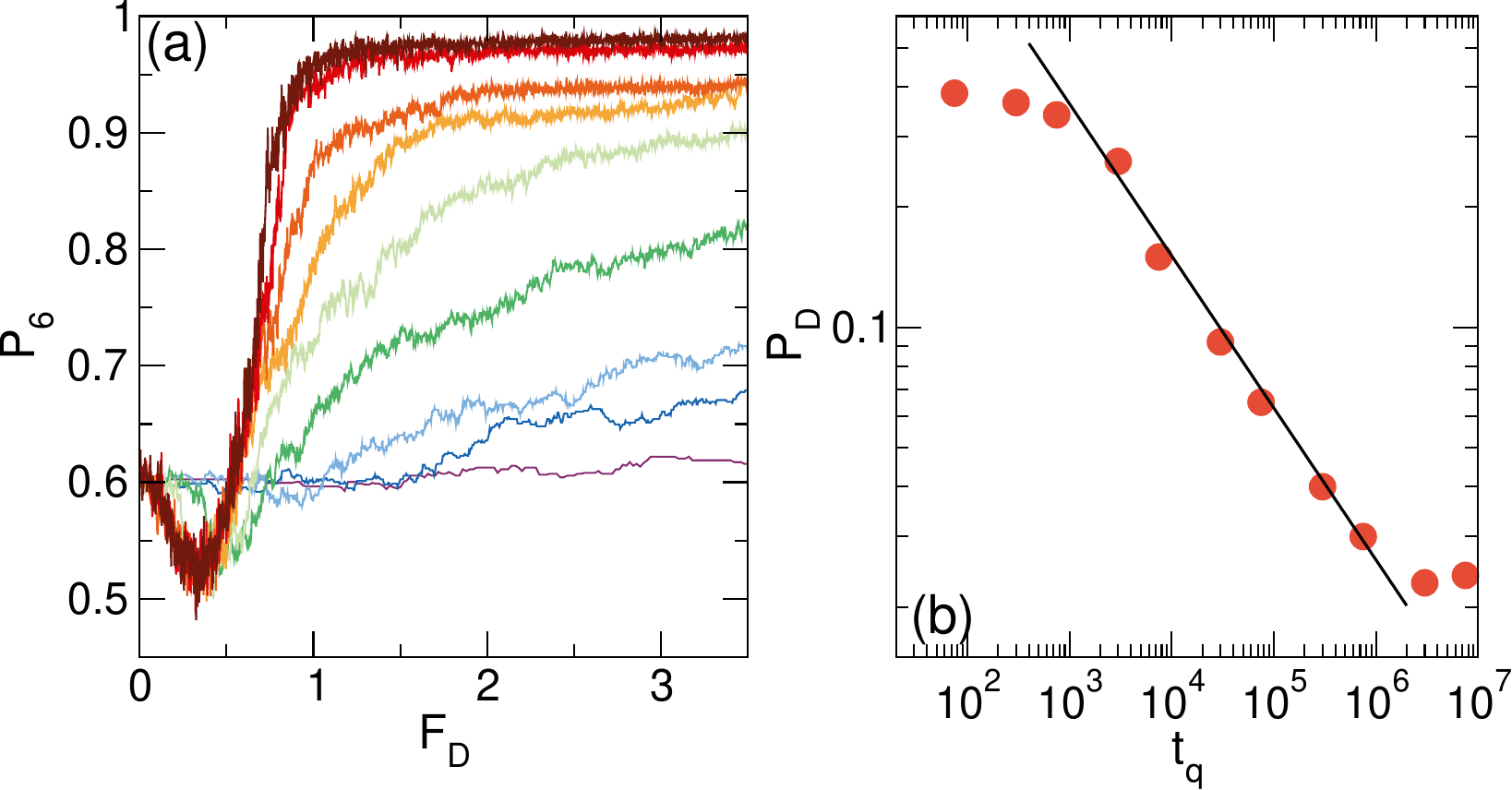}
\caption{
  {\bf Transition from the disordered to the ordered state
    for colloidal particles as a function
    of quenching speed.}
(a) $P_{6}$ versus $F_{D}$ for a colloidal system
with colloid density $n_{c} = 1.0$,
$n_{p} = 0.5$, $F_{p} = 1.0$, and $r_{p} = 0.35$
at $t_{q} = 3\times 10^6$, $7.5\times 10^5$, $3.0\times 10^5$,
$7.5\times 10^4$, $3.0\times 10^4$, $7.5\times 10^3$, 750, 300,
and 75, from top to bottom. 
(b) The defect density $P_{D}$ vs $t_{q}$ for the system in (a)
obtained at $F_{D} = 2.5$.
The solid line is a power-law fit with $\beta = -0.39$. 
}
\label{fig:5}
\end{figure}

In addition to superconducting vortices,
we have also considered colloidal particles driven over quenched disorder. 
Here we use the same type of overdamped simulations but
replace 
the particle-particle
interactions by the Yukawa or screened Coulomb
potential $U(r_{ij}) = A \exp(-\kappa r_{ij})/r_{ij}$ \cite{Reichhardt02},
with $\kappa=1.0$ and $A=0.1$.
The colloidal particles experience a
stronger short-range repulsion than the vortices.
In experiments, 
various types of random quenched disorder
can be introduced
along with an external driving force \cite{Tierno12a,Pertsinidis08},
and the number of topological defects can be
measured with imaging techniques \cite{Pertsinidis08,Deutschlander13}.
In Fig.~\ref{fig:5}(a) we show $P_{6}$ versus $F_{D}$ in a
colloidal system with colloidal density $n_{c} = 1.0$,
$n_{p} = 0.5$, $F_{p} = 1.0$, and $r_{p} = 0.35$
at $t_{q} = 3\times 10^6$, $7.5\times 10^5$, $3.0\times 10^5$,
$7.5\times 10^4$, $3.0\times 10^4$, $7.5\times 10^3$, 750, 300, and $75$,
where the final $P_{6}$ decreases for smaller $t_{q}$.
In Fig.~\ref{fig:5}(b) we plot
the final value of $P_{D}$ at $F_D=2.5$ versus $t_{q}$
for the system in Fig.~\ref{fig:5}(a).
The solid line is a power-law fit with $\beta = -0.39$,
similar to the exponent obtained for the vortex case.  

\section{Discussion}
We can ask whether the scaling exponent we obtain can be related
to possible universality classes of the dynamical transition. 
In the KZ scenario, the scaling exponent obeys
$\beta = (D-d)\nu/(1 + z\nu)$, where $\nu$ and $z$ are
critical exponents associated
with the universality class of the underlying phase transition,
$D$ is the spatial dimension of the system,
and $d$ is the dimension of the defect. For a 2D Ising model with
$D = 2$, 
$\beta = 2/3$, which is not what we observe;
however,
since the defects in our sample are all aligned in the
driving direction, our
system is closer to coupled 1D channels, which would give $\beta = 1/3$.
Recent simulations of certain quenched 2D spin ice models
give $\beta = 0.31$ \cite{Macauley20}.
A more likely candidate for the universality class of
a nonequilibrium system is directed percolation (DP)
\cite{Hinrichsen00}, where the critical exponents
depend on the effective dimension of the
dynamics. If we assume dynamics of dimension $1 + 1$,
we would have
$z = 1.58$ and $\nu = 1.097$ \cite{Hinrichsen00},
giving $\beta = 0.401$, which is in agreement with 
the values we observe.
We argue that the $1+1$ dynamics may be relevant
since the topological defects are mostly aligned in a single direction on
the ordered side of the transition, and
the relevant length scale could be the distance between the defects
in the direction of drive rather than perpendicular to the drive.

\begin{figure}
\includegraphics[width=3.5in]{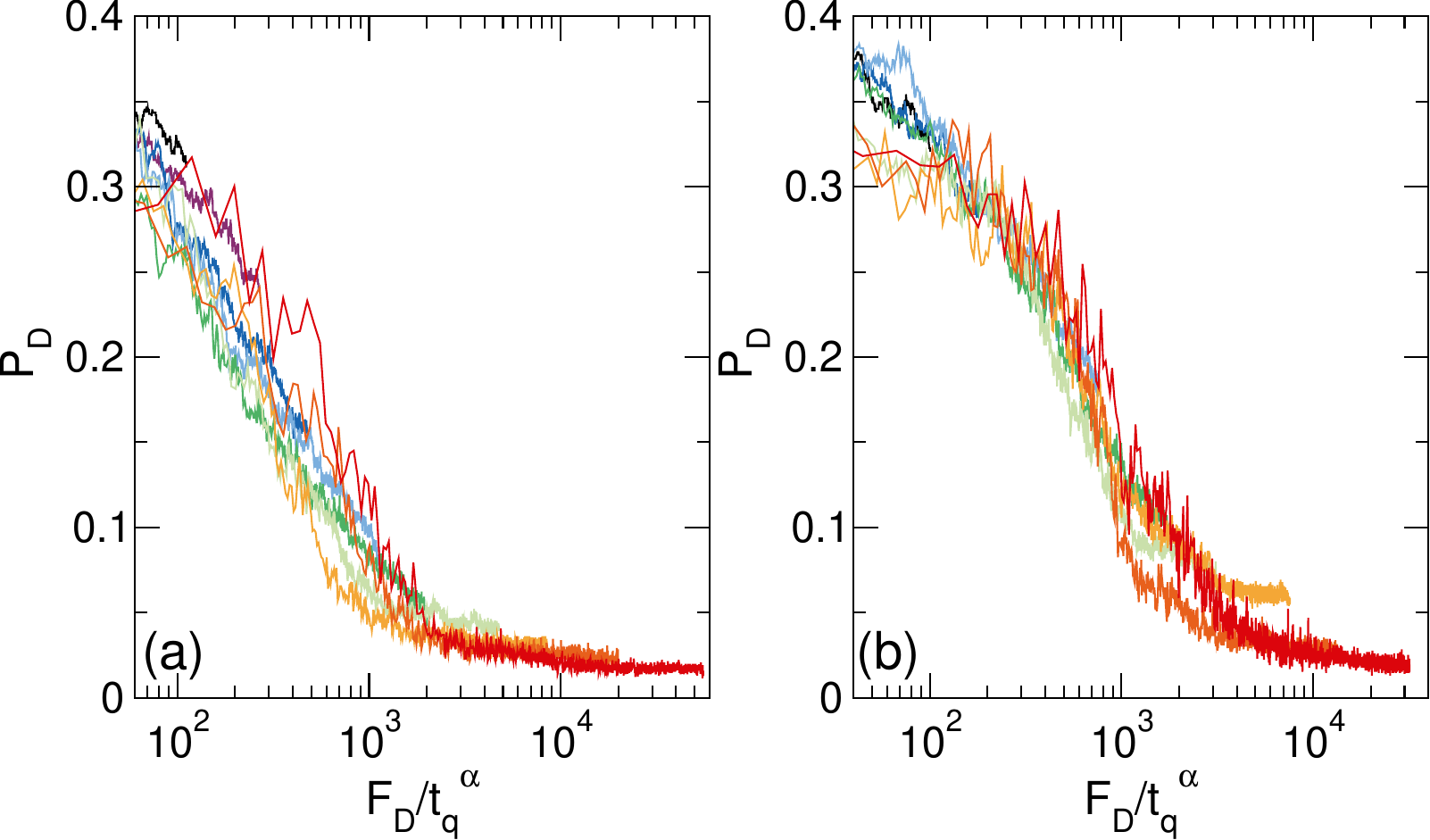}
\caption{
  {\bf Defect density scaling in the vortex and colloid systems.}
    Dimensionless
scaling plots of $P_{D}$ versus $F_{D}/t^\alpha_{q}$, where $t_{q}$ is the quench time
and where
$\alpha = 0.625$.
(a) The vortex system from Fig.~\ref{fig:3}(a).
(b) The colloidal system from Fig.~\ref{fig:5}(a).
}
\label{fig:6}
\end{figure}

In the KZ scenario, the lag time between the nonequilibrium and
equilibrium value is given by the freeze-out time
$\hat{t} \propto t_{q}^{z\nu}/(1 + z\nu)$ \cite{delCampo14}.
Since the value of the driving force $F_D$ is a function of time,
and since $F_D$ and $t_q$ are dimensionless,
this implies that the defect fraction should scale as
$P_{D} \propto F_{D}/t^\alpha_{q}$, where $t_{q}$ is the quench time. 
In Fig.~\ref{fig:6}(a) we illustrate this scaling
for the vortex system
from Fig.~\ref{fig:3}(a) and in Fig.~\ref{fig:6}(b) we show the scaling
for the colloidal system from Fig.~\ref{fig:5}(a). In each case,
the scaling exponent is
$\alpha = 0.625$. The KZ prediction gives $z\nu/(1 + z\nu) = \alpha$, where
plugging in
the scaling exponents for $1 + 1$ directed percolation \cite{Hinrichsen00}
leads to $\alpha = 0.632$, consistent with
the exponents we find. 

\begin{figure}
\includegraphics[width=3.5in]{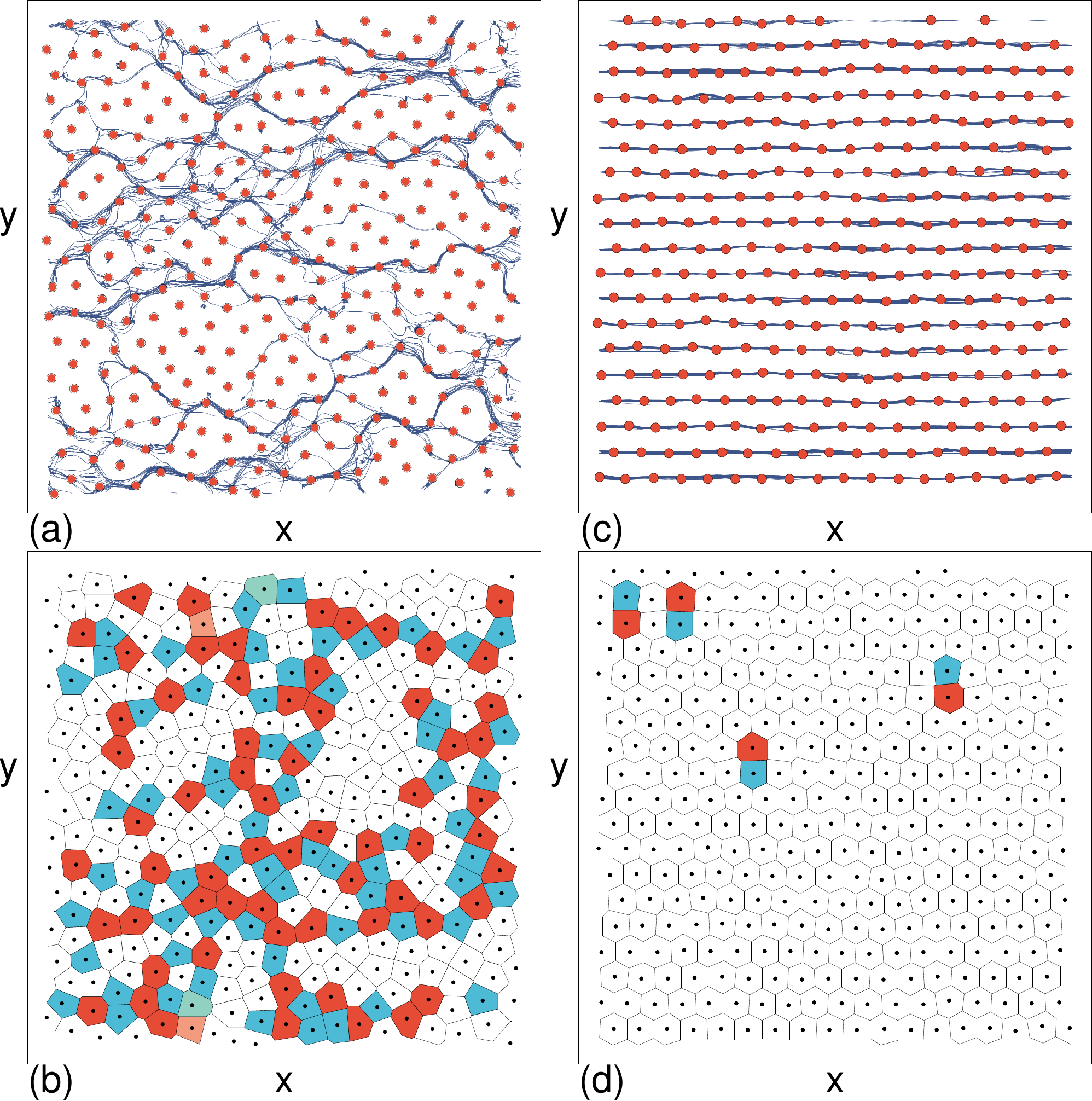}
\caption{
  {\bf Disordered and 1+1 dimensional flow
    below and above the dynamical phase transition.}
  (a) The colloid positions (red circles) and trajectories (lines)
for a subset of the system in Fig.~\ref{fig:5}
with $n_c=1.0$, $n_p=0.5$, $F_p=1.0$, and $r_p=0.35$
at $t_{q} = 7.5\times 10^5$ and $F_{D} = 0.25$, showing
2D disordered flow.
(b) The corresponding Voronoi plot showing a high defect density.
(c) The colloid positions and trajectories in the same system
at $F_{D} = 1.5$ showing 1D channeling.
(d) The corresponding Voronoi plot
showing that all of the
topological defects are aligned in the same direction. 
}
\label{fig:7}
\end{figure}

To better understand why the dynamics is $1 + 1$ dimensional,
in Fig.~\ref{fig:7}(a) we illustrate the colloidal positions
and trajectories on the
disordered side of the transition for a subset of the
system in Fig.~\ref{fig:5} at $F_{D} = 0.25$ and $t_{Q} = 7.5\times10^5$,
showing strongly disordered flow occurring in both the $x$ and $y$ directions.
The corresponding Voronoi plot in Fig.~\ref{fig:7}(b)
contains a high defect density, similar to what is observed
in the vortex system in the disordered phase.
Figure~\ref{fig:7}(c) shows the positions and trajectories of the
colloids on the ordered
side of the transition at $F_{D} = 1.5$.
Here the dynamics
are strictly 1D in character, and
the topological defects are all aligned
in the direction of drive as shown in Fig.~\ref{fig:7}(d).
In theoretical work for particles
moving over random disorder in 2D,
it is argued that the strongly driven case
can be considered as a series of coupled 1D channels that
slide past one another to form a moving smectic state
\cite{Balents98,LeDoussal98}. This 1D channeling could explain why
the dynamics produce scaling exponents consistent with
$1+1$ DP rather than $2 + 1$ DP.  

In conclusion, we have investigated the evolution of the density of defects
across a dynamical disorder to order nonequilibrium phase transition for
driven particles moving over quenched disorder
as we vary the quench rate.
We find that the defect density scales as a power law
with quench rate, in agreement with the predictions
of the Kibble-Zurek scenario. 
For both superconducting vortices and colloidal assemblies,
we find a scaling exponent of $\beta \approx -0.39$,
which is consistent with an underlying transition
that falls 
in the 1+ 1 directed percolation universality class since
the ordered system
forms a moving smectic state in which
the defects are aligned in 1D chains.  
Experimentally, our predictions could be tested
in superconducting vortex systems using
various imaging and
transport measures that have previously been shown to be
correlated with the number of defects in the
vortex lattice \cite{Hellerqvist96}.
Direct imaging of the dynamics is feasible using colloidal systems,
and it would also be possible to consider ac drives
which
would avoid edge effects. Our results
suggest that the Kibble-Zurek scenario 
can be applied to more general non-equilibrium continuous
phase transitions, particularly those 
between disordered and ordered states. Our results also imply that
a system undergoing
a directed percolation transition could exhibit features of
the Kibble-Zurek scenario provided that
some type of well-defined defect structure can be identified. 

 \smallskip
 
\noindent{\textbf{\textsf{Acknowledgements}}}\\ 
We thank W. H. Zurek for useful discussions.
We gratefully acknowledge the support of the U.S. Department of
Energy through the LANL/LDRD program for this work.
This work was supported by the US Department of Energy through
the Los Alamos National Laboratory. Los Alamos National Laboratory is
operated by Triad National Security, LLC, for the National Nuclear Security
Administration of the U. S. Department of Energy (Contract No. 892333218NCA000001).

\medskip

\noindent{\textbf{\textsf{Competing interests}}}\\
 The authors declare no competing interests.
 
 \smallskip

\noindent{\textbf{\textsf{Data availability}}}\\
The datasets generated during and/or analyzed during
the current study are available from the corresponding
author upon reasonable request.

\smallskip

\noindent{\textbf{\textsf{Code availability}}}\\
The codes generated and used during the current study
are available from the corresponding author on reason-
able request.

 \smallskip
 

\smallskip

\noindent{\textbf{\textsf{Author contributions}}}\\
 CJOR and CR conceived and led the development of the project, performing the numerical analysis.
All the authors participated in the analysis of the results and the writing of the manuscript.

\bibliographystyle{naturemag}

\bibliography{mybib}

\end{document}